\documentclass[12pt,preprint]{aastex}

\slugcomment{to be submitted to the Astrophysical Journal}

\shorttitle{
The White Dwarf in SS Cygni}
\shortauthors{Sion et al.}

\begin{document}

\title{The Accreting White Dwarf in SS Cygni Revealed\altaffilmark{1}} 

\author{Edward M. Sion, Patrick Godon\altaffilmark{2}, Janine Myzcka} 
\affil{Astronomy \& Astrophysics, Villanova University, \\ 
800 Lancaster Avenue, Villanova, PA 19085, USA}
\email{
edward.sion@villanova.edu, 
patrick.godon@villanova.edu, 
janine.myzcka.villanova.edu}

\and 

\author{William P. Blair}
\affil{Department of Physics and Astronomy, The Johns Hopkins Univeristy, \\  
3400 N. Charles Street, Baltimore, MD 21218}
\email{wpb@pha.jhu.edu}

\altaffiltext{1}{
Based on observations made with the NASA/ESA Hubble Space Telescope,
obtained at the Space Telescope Science Institute, which is operated by the
Association of Universities for Research in Astronomy, Inc. under NASA
contract NAS5-26555, and the NASA-CNES-CSA {\it Far Ultraviolet Spectroscopic 
Explorer}, which is operated for NASA by the Johns Hopkins University under
NASA contract NAS5-32985.
}
\altaffiltext{2}{Visiting at the Johns Hopkins University,
Henry A. Rowland Department of Physics and Astronomy, 
Baltimore, MD 21218}

\clearpage 

\begin{abstract} 

We have carried out a combined Hubble Space Telescope (HST/GHRS) and
Far Ultraviolet Spectroscopic Explorer ({\it FUSE})  
analysis of the prototype dwarf nova SS Cygni during quiescence. 
The {\it FUSE} and HST spectra were obtained at comparable times after 
outburst and have matching flux levels where the two spectra overlap. 
In our synthetic spectral analysis, we have used SS Cygni's accurate 
HST FGS parallax giving $d = 166 $ pc, a newly determined mass for the 
accreting white dwarf 
\citep{bit07}  
of $M_{wd} = 0.81 M_{\odot}$ (lower than the previous, widely used 
$1.2 M_{\odot}$) and 
the reddening ($E_{B-V}$) values 0.04 \citep{ver87,lad91} \& 0.07 \citep{bru94} derived from
the 2175 \AA\ absorption feature in the IUE LWP spectra. From the best-fit model solutions  
to the combined HST + {\it FUSE} spectral energy distribution, we find that
the white dwarf is reaching a temperature $T_{eff} \approx 45-55,000$ K in quiescence, assuming 
$\log(g)= 8.3$ with a solar composition accreted atmosphere. The exact temperature
of the  WD depends on the reddening assumed and on the inclusion of a quiescent low mass
accretion rate accretion disk.   
Accretion disk models alone  fit badly in the {\it FUSE} range while, and if we take the distance 
to be a free parameter, the only accretion disk model which fits well 
is for a discordant distance of at least 
several hundred pc and an accretion rate 
($\sim 10^{-8} M_{\odot}$/yr) which is unacceptably high for a dwarf nova in 
quiescence. We discuss the implications of the white dwarf's 
temperature on the time-averaged accretion rate and long term 
compressional heating models.

\end{abstract} 

 
\clearpage 

\section{Introduction} 

SS Cygni is a prototype dwarf nova, a subclass of the cataclysmic 
variable group of binary stars. Cataclysmic variables contain a 
white dwarf and a larger radius, less massive, donor star. The donor star 
fills its Roche lobe and its matter has gradually been cannibalized 
by the white dwarf. In dwarf novae the matter feeds into a disk 
around the white dwarf and continues to build up the disk mass until 
the disk reaches a critical temperature at which time the mass 
collapses onto the white dwarf and releases gravitational potential 
energy as radiation. 
This results in a roughly periodic increase in brightness of the system. 

SS Cyg is one of the best-studied dwarf novae. It has an 
orbital period of 6.6 hr (well above the period gap) and a $\sim 50$ day 
recurrence time between dwarf nova outbursts 
\citep{can92,can93}. 
These factors and its 
well-studied history have made it a key target of our ongoing studies 
to understand how disk accretion affects the white dwarf. Observations 
must be taken when the system is in quiescence because in such a state, 
the disk is least luminous which offers a favorable opportunity to detect 
the radiation of the white dwarf photosphere. However, the source of far 
ultraviolet light during the quiescence of dwarf novae remains controversial. 
This is especially true in dwarf novae with orbital periods above the CV 
period gap, which tend to have higher mass transfer rates, larger accretion 
disks and more massive secondary stars. Therefore, during quiescence it 
is possible that some portions of the remaining disk may still  be optically 
thick and dominate the far ultraviolet (FUV) flux of the system.

The inclination of its orbit is $\sim 51^{\circ}$ \citep{bit07}
while its white dwarf mass 
has been reported     to be $1.2 M_{\odot}$ \citep{sha83} and 
secondary mass $0.7 M_{\odot}$. The system has an accurate parallax 
measured with the 
Hubble Space Telescope fine guidance sensor
\citep{har00}  
yielding a parallax distance of $166^{+14}_{-12}$ pc.
\citet{hol88} 
first noted the possible detection of the underlying 
accreting white dwarf from the occasional appearence of Ly$\alpha$  
absorption in some IUE spectra obtained during quiescence.
However, despite a number of attempts
to measure the white dwarf's temperature and quantify its contribution to 
the FUV flux, the uncertain contribution of the accretion disk and 
other possible unidentified source(s) of FUV emission, the "second component", 
rendered any such measurements suspect 
\citep{hol88b,les03,lon05}.

Recently however, we retrieved 
archival {\it FUSE} and HST spectra, both obtained 
during different quiescent intervals of SS Cygni but with closely matching 
flux levels in the wavelength range where HST/GHRS and {\it FUSE} overlap 
($\sim 1150-1190$ \AA\ ). 
Also a new mass determination for the white dwarf 
has now been derived by \citet{bit07}:  
$0.81 M_{\odot} \pm0.18$, 
significantly lower than the earlier value of $1.2 M_{\odot}$. 
Using this lower mass and the parallax 
distance of 166 pc, we have reexamined the nature of the hot component 
with synthetic spectral models of high gravity photospheres and optically 
thick accretion disks. This assessment is presented below.  

In the next section we give details of the archival spectra, in section 3
we present our spectal modeling tools and  method, results are presented in
section 4 and discussed in the concluding section.   

\section{Archival Observations}

The archival HST GHRS spectrum was taken 
on JD 2450352 (1996 September 26), in mid-quiescence, 17 days after SS Cygni 
reached quiescence and 21 days after the previous outburst. The HST 
spectrum is a combination of two individual spectral segments  
Z3DV0204T (1150 \AA - 1435 \AA ) and  
Z3DV0205T (1377 \AA - 1663 \AA )  
taken in ACCUM mode with the GHRS spectrograph 
with the G140L grating and LSA aperture (of size 1.74").   
The two spectral segments, one at each grating setting, 
had an exposure time of 2176s each,  
and there was a 32.25 second time gap in an otherwise 
nominal exposure. The GHRS observations were calibrated using the 
standard pipeline CALHRS, and were retrieved as VO-Tables using
VOSpec.  

The archival {\it FUSE} spectrum was taken on JD 2452156.5 (2001 September 4),  
18 days after the previous outburst, and 13 days after the 
system first entered quiescence. 
The {\it FUSE} spectrum (P2420101), was a combined spectrum 
of 8 separate exposures taken through the 30"x30" LWRS Large Square Aperture 
in TIME Tag mode. The total 
good exposure time was $\sim 18$ks, varying slightly for each spectral channel.   
The observations were carried out mostly during NIGHT time. 
The {\it FUSE} observations were processed with CalFUSE version v3.2.3
\citep{dix07}. 
We follow the same procedure we used previously for the postprocessing 
(co-addition, alignment and weight of the spectral channels) of the {\it FUSE} 
data (see e.g. \citet{god06}).  

The combined {\it FUSE} + GHRS spectrum of SS Cyg is presented in
Fig.1 (see section 4 for the model fit).  
The spectrum is characterized by strong emission lines originating
in an optically thin region in the disk or above it, as inferred 
from their rotational broadening (see \citet{lon05} for a rigourous
analysis of these lines in the GHRS spectra).  In the very short
wavelengths of {\it FUSE} the broad emission lines from 
N\,{\sc iv}, S\,{\sc vi} and H\,{\sc i} (Ly$\delta$ and higher) 
merge together and produced an apparent rise of flux 
($< 950 $ \AA ), which we do not attempt to model. In the
fitting (section 4) we mask all the emission lines and these
appear in blue in Fig.1. It is likely that around 1060-1080 \AA\  
and 1110-1120 \AA\ some  additional emission is present.

\section{Synthetic Spectral Modeling} 

We adopted model accretion disks from the optically thick disk model grid 
of \citet{wad98}. In these accretion disk models, the innermost 
disk radius,
$R_{in}$, is fixed at a fractional white dwarf radius
of $x = R_{in}/R_{wd} = 1.05$. The outermost disk radius, R$_{out}$, was chosen
so that $T_{eff}(R_{out})$ is near 10,000 K since disk annuli beyond this point, 
which are cooler zones with larger radii, would provide only 
a very small contribution to the 
mid and far UV disk flux, particularly the FUV bandpass ($\sim 900-1700$ \AA ). 
The mass transfer rate is 
assumed to be the same for all radii.

Theoretical, high gravity, photospheric spectra were computed by first 
using the code
TLUSTY \citep{hub88} to calculate the atmospheric structure and SYNSPEC 
\citep{hub95}                  
to construct synthetic spectra. We compiled a library of photospheric spectra  
covering the temperature range from 15,000 K to 70,000 K in increments 
of 1000 K, and a surface gravity range, 
$\log(g)= 7.0 - 9.0$, in increments of 0.2 in $\log(g)$.

The reddening of the system was taken from estimates 
listed in the literature, determined from the 
strength of the 2200 \AA\ interstellar absorption feature.  
\citet{ver87} and \citet{lad91} both give E(B-V)=0.04 while the more recent
work of \citet{bru94} gives E(B-V)=0.07. Both values are much
smaller than the galactic reddening in the direction of SS Cyg
which is pretty large ($\sim 0.5$) in agreement with the fact that
SS Cyg is rather nearby with a distance of {\it only} 166 pc.  
The combined spectrum was  de-reddened with the IUERDAF IDL routine UNRED 
assuming both E(B-V)=0.04 and E(B-V)=0.07.

After masking emission lines in the spectra, we 
determined separately for each spectrum, the best-fitting white dwarf-only 
models and the best-fitting disk-only models using  a $\chi^{2}$ 
minimization routine. A $\chi^{2}$ value and a scale factor were 
computed for each model fit. The scale factor, $S$, normalized to a kiloparsec 
and solar radius, can be related to the white dwarf radius R through: 
$$  
F_{\lambda(obs)} = S H_{\lambda(model)}, ~~~~~~~~~~~where ~~~~S=4\pi R^{2} d^{-2}, 
$$  
and $d$ is the distance to the source. For the white dwarf radii, we use 
the mass-radius relation from the evolutionary model grid of 
\citet{woo95} 
for C-O cores. We combined white dwarf models and accretion disk models 
using a $\chi^{2}$ minimization routine called DISKFIT. Using this method 
the best-fitting composite white dwarf plus disk model is determined based 
upon the minimum $\chi^{2}$ value achieved, visual inspection of 
the model, consistence with the continuum slope and Ly$\alpha$  region, 
and consistence of the scale factor-derived distance with the adopted 
trigonometric parallax distance.

\section{Synthetic Spectral Fitting Results}

We used an accretion disk model, a white dwarf photosphere, and a 
combination of both in our analysis.  
The disk models used were optically thick, steady state accretion 
models with solar abundances.  They are considered a reasonable 
first approximation to the spectral shape of the disk in quiescence. 
In the disk models, we first adopted an inclination of 
$41^{\circ}$ and  $60^{\circ}$  directly from the grid of models of  
\citet{wad98}. Only after a satisfactory best fit was obtained with the 
correct (parallax) distance, did we then  
generate disk models 
(using TLUSTY, SYNSPEC and DISKSYN) 
with an inclination of $50^{\circ}$,  
closer to its derived value of $51^{\circ} \pm 5$ \citep{bit07}. 
The photosphere models were generated using TLUSTY and SYNSPEC.
In the following we used the parallax distance of 166 pc, and the 
mass of $0.8 M_{\odot}$ ($\log(g)= 8.3$)  \citep{bit07}.  
We assumed reddening values of E(B-V) = 0.04 and 0.07. 
A summary of the model fits is given in Table 1.

We started our fitting with single accretion disk models alone 
assuming both E(B-V)=0.04 \& 0.07.  
We ran disk model fits and found that the best fit (lowest $\chi^2$)
had a mass accretion rate far too large for quiescence and a distance
much larger than 166 pc.

Next, we tried single white dwarf atmosphere models, first dereddening
assuming E(B-V)=0.04.  
Since almost all the lines are in emission (possibly from an optically
thin region in the disk and/or corona), it makes it 
difficult to assess the rotational velocity broadening and chemical
abundances  based on the absorption lines.  
Nevertheless, for all WD models we assumed solar abundances and 
a canonical projected rotational 
velocity of 200 km/s, and checked that the value of the $\chi^2$
did not depend on $V_{rot}\sin{i}$ as long as it was a few hundred
km/s.  
The best fit least $\chi^2$ WD model has a temperature
$T_{eff}=40,000$ K and $\chi^2=1.637$, but with a distance of only 138 pc. 
A model with $T_{eff}=47,000$ K gave the right distance with a slightly
larger $\chi^2$, namely 1.990.  
Next we ran single WD model fits assuming E(B-V)=0.07. The $\chi^2$
we obtained increased slightly over the E(B-V)=0.04 best fit models. 

Last, we explored whether the  fitting could be improved if we 
combined a white dwarf model with an accretion disk model. 
We found that some of the white dwarf plus disk combinations yielded 
distances close to 166 pc with a lower $\chi^2$.
We have summarized some of these combination fits in Table 1.
For E(B-V)=0.04, the best WD+disk fit leading to a distance 
in agreement with observed parallax is for a WD with 
$T=46,000$ K, a disk with a mass accretion rate 
$\dot{M}=1 \times 10^{-10} M_{\odot}$/yr, $i=50^{\circ}$, 
where the WD contributes
88\% of the flux and the disk contributes the remaining 12\%.
This model fit is presented in Figure 1.  
We then carried out the same fitting but this time assuming 
E(B-V)=0.07 and found similar results for the disk but with a higher
WD temperature: $T_{eff}=55,000$ K.  

Therefore, we conclude 
that the dominant source of the far UV radiation
between 912 \AA\ and 1660 \AA\ is an accretion-heated white dwarf photosphere with 
$T_{eff} \approx 45-55,000$ K, $\log(g)= 8.3$ (depending on the 
reddening value).

\section{Conclusions} 

We have presented evidence from our model fitting analysis of the combined 
{\it FUSE} + HST/GHRS spectrum of SS Cyg during quiescence that
the source of the far ultraviolet continuum and absorption line radiation 
is the white dwarf's photosphere. The disk models that best fit the spectral 
data yield unreasonably large distances, at odds with the HST FGS parallax 
and indicate accretion rates
far too high to be associated with dwarf nova quiescence.
The photosphere models give effective temperatures of 45,000 to 55,000 K 
for a reddening of 0.04 and 0.07 with the inclusion of a low mass accretion
rate disk in agreement with the quiescent state. Unfortunately, the lack of a clear
detection of absorption lines
due to accreted metals and helium in the WD atmosphere prevents a determination of both 
the rotational velocity of the white dwarf and the abundance of metals 
in its accreted atmosphere.

Our derived temperature for the white dwarf in SS Cygni is well above 
the presently estimated average temperature ($\sim$30,000 K)
for white dwarfs in dwarf novae above the CV period gap. 
Compared with the temperatures of white dwarfs in dwarf novae whose 
orbital periods are close to the period of SS Cygni, the white dwarf in 
TT Crt is cooler (40,000 K; 
\citep{sio08}) 
while the white dwarf in 
Z Cam is hotter (57,000 K; 
\citep{har05}). 
If the accretion rate 
scales with the orbital period, then the temperatures should be comparable. 
It is interesting that the white dwarfs in dwarf novae with
$ P_{orb} <  360$ mins are cooler than 40,000 K while the  hottest 
white dwarfs in dwarf novae are found at $P_{orb} > 360$ mins.

At $P_{orb} = 6.6$ h, the $T_{eff}$ of the WD in SS Cygni lies within the 
range expected from compressional heating for an average 
$\dot{M}$, $<\dot{M}>$, obtained from 
typical interrupted magnetic braking laws for white dwarf masses between
$0.6 M_{\odot}$ and $1.0 M_{\odot}$ \citep{tow09}.  
A linear extrapolation to $P_{orb}$, of 
the predicted $T_{eff}$ for $P_{orb} = 6.6$ h, 
corresponds to an average mass transfer rate
of $<\dot{M}> \sim  10^{-8} M_{\odot}$/yr 
which is at the high end of the range of 
mass transfer rates associated with the nova-like 
variables, as determined from their 
optical disk luminosities \citep{war95}.   
Interestingly, the braking laws of \citet{and03}  
and \citet{iva04}  
either fall drastically short or exhibit a downturn, respectively, 
of yielding the $<\dot{M}>$ 
implied by the WD $T_{eff}$, while the \citet{how01}  
law steeply increases at constant $P_{orb} \sim  5$ h.
One cannot rule that other significant sources of 
heating of the white dwarf besides compression are operating
such as possible nuclear burning. It seems clear that more white dwarf 
temperatures are needed in dwarf novae and nova-like
variables at long $P_{orb}$ before definitive conclusions can be drawn.

\section{acknowledgements} 
We thank an anonymous referee for useful comments.
PG thanks the Henry A. Rowland Department of Physics and Astronomy 
at the Johns Hopkins University for hospitality. 
This work was supported in part by NSF grant AST-0807892 to Villanova University,
a Villanova Undergraduate Research Grant. 
Support for this work was also provided by NASA through grant number HST-AT-10657.01
to Villanova University from the Space Telescope Science Institute, which is
operated by the Association of Universities for Research in Astronomy, Incorporated, 
under NASA contract NAS5-26555. Additional support was provided by the National
Aeronautics and Space Administration (NASA) under Grant number NNX08AJ39G issued
through the office of Astrophysics Data Analysis Program (ADP) to Villanova University.

\begin{deluxetable}{cccccccc}

\tablewidth{0pt}
\tablecaption{Synthetic Spectral Model Fits}
\tablehead{
$Log(\dot{M})$  & $i$  & $T_{eff}$(WD) & $\chi^2$ &  $d_{model}$ &    WD(\%)  &  disk(\%) & E(B-V)    \\ 
$<M_{\odot}/yr>$ & $<deg>$ & $<K>$     &             & $<pc>$    &            &           &               
}
\startdata
-8.0  & 41    &    ---     &    1.331 &    862   &    ---  &    100   & 0.04    \\ 
-8.0  & 60    &    ---     &    1.227 &    629   &    ---  &    100   & 0.04    \\ 
-9.0  & 41    &    ---     &    1.989 &    308   &    ---  &    100   & 0.04    \\ 
-9.0  & 60    &    ---     &    2.690 &    216   &    ---  &    100   & 0.04    \\ 
-9.5  & 41    &    ---     &    6.477 &    157   &    ---  &    100   & 0.04    \\ 
-9.5  & 50    &    ---     &    8.036 &    142   &    ---  &    100   & 0.04    \\ 
-9.5  & 60    &    ---     &    9.122 &    106   &    ---  &    100   & 0.04    \\ 
-8.0  & 41    &    ---     &    1.615 &    741   &    ---  &    100   & 0.07    \\ 
-8.0  & 60    &    ---     &    1.810 &    541   &    ---  &    100   & 0.07    \\ 
-9.0  & 41    &    ---     &    3.562 &    265   &    ---  &    100   & 0.07    \\ 
-9.0  & 60    &    ---     &    4.754 &    186   &    ---  &    100   & 0.07     \\ 
 ---  & ---   &    40,000  &    1.637 &    138   &    100  &    ---   & 0.04    \\ 
 ---  & ---   &    47,000  &    1.990 &    167   &    100  &    ---   & 0.04     \\ 
 ---  & ---   &    46,000  &    1.451 &    139   &    100  &    ---   & 0.07    \\ 
 ---  & ---   &    55,000  &    1.600 &    164   &    100  &    ---   & 0.07    \\ 
-10.5 &   50  &    41,000  &    1.490 &     143  &   97.6  &   2.4    & 0.04    \\ 
-10   &   50  &    46,000  &    1.258 &     173  &   88.0  &  12.0    & 0.04     \\ 
-9.5  &   50  &    55,000  &    1.255 &     233  &   66.3  &  33.7    & 0.04    \\ 
-10.5 &   50  &    49,000  &    1.429 &     149  &   98.4  &   1.6    & 0.07    \\  
-10   &   50  &    55,000  &    1.385 &     172  &   91.0  &   9.0    & 0.07    \\  
-9.5  &   50  &    70,000  &    1.630 &     222  &   72.4  &  27.6    & 0.07    \\

\enddata
\end{deluxetable}

\clearpage

\begin{figure}
\vspace{-3.cm} 
\plotone{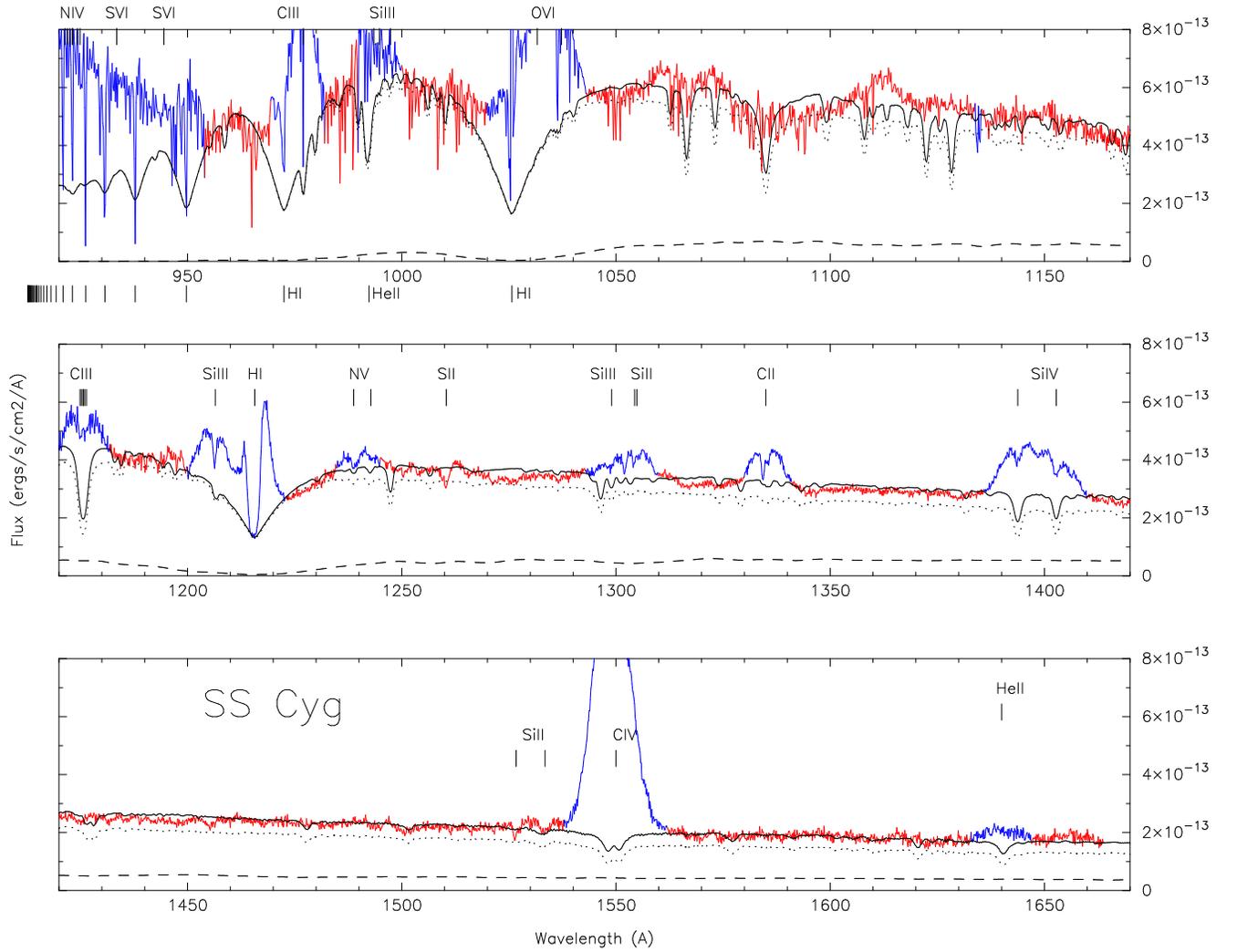}
\figurenum{4}
\caption{The combined spectrum of SS Cyg (in red) dereddened assuming E(B-V)=0.04 
is shown with a composite  WD+disk  synthetic spectrum (in solid black). 
The WD model (dotted line) has  
$M_{wd} = 0.8 M_{\odot}$ (corresponding to $Log(g)=8.3$), 
$T_{eff}=46,000$ K, solar composition and a projected rotational velocity
of 200km/s. The dashed line is the contribution from the accretion disk model.
The disk model has a mass accretion rate $\dot{M}=10^{-10}M_{\odot}$/yr and an  inclination $i=50^{\circ}$. 
The $\chi^2$ obtained is 1.258 and the resulting distance is 173 pc.
The WD contributes 88\% of the FUV flux, while the disk contributes the remaining 12\%.  
The excess of flux around 1060-1080 \AA\ , and 1110-1130 \AA\   might be due to some emission
from 
S\,{\sc iv} (1063, $\sim$1073 \AA ),  
Si\,{\sc iv} (1066\AA ),  
Si\,{\sc iii} ($\sim$1108-1113 \AA ),  
and 
Si\,{\sc iv} (1122.5, 1128.3 \AA ).  
}
\end{figure}

\end{document}